\documentclass[12pt]{article}
\usepackage{enumerate}
\usepackage{amsfonts}
\usepackage{amsmath}
\usepackage{amssymb}
\usepackage{amsthm}

\def\H{\mathcal{H}}
\def\K{\mathcal{K}}
\def\P{\mathfrak{P}}
\def\S{\mathfrak{S}}
\def\T{\mathfrak{T}}

\newcommand{\Tr}{\mathrm{Tr}}

\newcounter{defin}  \newcounter{lemma}  \newcounter{theorem}
\newcounter{property} \newcounter{corol}  \newcounter{remark} \newcounter{example}

\newenvironment{property}{\par\refstepcounter{property}
     \textbf{Proposition \theproperty.}\ }{\rm\par}

\begin{document}
\title{On a conjectured property of the von Neumann entropy valid in the commutative case.}
\author{M.E. Shirokov\footnote{email:msh@mi.ras.ru}\\
Steklov Mathematical Institute, RAS, Moscow}
\date{}
\maketitle \vspace{-15pt}
\begin{abstract}
It is well known that the von Neumann entropy is continuous on a subset of quantum states with bounded energy provided the Hamiltonian $H$ of the system satisfies the condition $\Tr\exp(-cH)<+\infty$ for any $c>0$. In this note we consider the following conjecture: every closed convex subset of quantum states, on which the von Neumann entropy is continuous, consists of states  with bounded energy with respect to a particular Hamiltonian $H$ satisfying the above condition.

It is shown that the classical analog of this conjecture is valid (i.e. it is valid for the Shannon entropy). It is also shown that this conjecture holds for some types of subsets consisting of non-commuting states, but its validity for all subsets of quantum states remains an open question.
\end{abstract}
\maketitle


\section{Preliminary notes}

Let $\mathcal{H}$ be a separable Hilbert space and $\S(\H)$ be the
set of quantum states -- positive operators in $\H$ with unit trace. The von
Neumann entropy of a state $\rho\in\S(\H)$ is defined by the formula
$H(\rho)=-\Tr\rho\log\rho$. The function $\rho\mapsto H(\rho)$ is
nonnegative concave and lower semicontinuous on the set $\S(\H)$
taking the value $+\infty$ on a dense subset of $\S(\H)$
\cite{O&P,W}. In analysis of infinite dimensional quantum systems
and channels different continuity conditions for the von Neumann
entropy are used \cite{S,W,Sh-4,Sh-11} (see the overview in \cite{Sh-11}).

Among all possible subsets of $\S(\H)$ the closed convex subsets
$$
\mathcal{K}_{H,h}=\{\rho\in\S(\H)\,|\,\Tr H\rho\leq h\}
$$
(defined by a positive operator $H$ in $\H$ and $h>0$) play a special role.
If $H$ is a Hamiltonian of a quantum system associated with the
space $\H$ then $\mathcal{K}_{H,h}$ is the set of all quantum states with
the mean energy not exceeding $h$.

One of the most useful continuity conditions for the von Neumann
entropy (originally appeared in \cite{W} as far as I know) states
that the function $\rho\mapsto H(\rho)$ is continuous on the set
$\mathcal{K}_{H,h}$ if $\Tr\exp(-\lambda H)<+\infty$ for all $\lambda>0$, i.e. if
$$
\mathrm{g}(H)=\inf\{\lambda>0\,|\,\mathrm{Tr}\exp(-\lambda
H)<+\infty\}=0.
$$

A detailed analysis of a restriction of the von Neumann entropy to
the set $\mathcal{K}_{H,h}$ is made in \cite[Proposition 1]{Sh-4},
where it is shown, in particular, that the condition
$\mathrm{g}(H)=0$ is necessary for continuity of the function
$\rho\mapsto H(\rho)$ on the set $\mathcal{K}_{H,h}$ and that this function
is finite on the set $\mathcal{K}_{H,h}$ if and
only if $\mathrm{g}(H)<+\infty$. It is also shown that any convex
closed set of states, on which the von Neumann entropy is finite, is
contained in the set $\mathcal{K}_{H,h}$ for particular operator $H$ with
$\mathrm{g}(H)<+\infty$ and $h>0$ \cite[Corollary 5]{Sh-4}.\footnote{Finiteness of the entropy on a convex
closed set of states implies its boundedness.} The last observation is quite useful, in particular, it plays a basic role in the proof of equivalence of (global) finiteness and continuity of the output entropy of a positive map \cite[Theorem 1]{Sh-12}. So, we have the following
collections of statements\medskip
$$
\begin{array}{ccc}
 H(\rho)\; \text{is finite  on the set}\;
\mathcal{K}_{H,h}\quad &\Leftrightarrow& \quad g(H)<+\infty,\quad
\\\\
 H(\rho)\; \text{is continuous on the set}\;
\mathcal{K}_{H,h}\quad &\Leftrightarrow &\quad
g(H)=0,\quad\\\\
H(\rho)\; \text{is finite on a  closed convex set}\; \mathcal{A}
&\Leftrightarrow&
 \exists H,g(H)<+\infty\, :\;
 \mathcal{A}\subseteq\mathcal{K}_{H,h}.
\end{array}\vspace{5pt}
$$
Simple examples show that the conditions of closedness and convexity
of the set $\mathcal{A}$ in the last statement is essential (for the
implication $"\Rightarrow"$).\smallskip

To make the above collections of statements complete it is
reasonable to conjecture validity of the following statement
$$
H(\rho)\; \text{is continuous on a  closed convex set}\; \mathcal{A}
\;\,\Leftrightarrow\;\, \exists H,g(H)=0\,:\,
\mathcal{A}\subseteq\mathcal{K}_{H,h}.\vspace{5pt}
$$
Note again that the implication $"\Rightarrow"$ in this statement does
not hold if either the set $\mathcal{A}$ is not closed or is not
convex. Indeed, to show necessity of the convexity condition one can
take the sequence $\lambda_n|n\rangle\langle n|+(1-\lambda_n)|1\rangle\langle 1|$, where $\lambda_n\rightarrow 0$ and $\{|n\rangle\}$ is a basis
in $\H$,  necessity of the closedness condition follows
from existence of convex sets $\mathcal{A}$ such that the von
Neumann entropy is continuous on $\mathcal{A}$ but is not continuous
on its closure $\mathrm{cl}(\mathcal{A})$ (see the example in
\cite[Remark 5]{Sh-11}).

\section{The classical case}

In this section we will show validity of the above-stated conjecture
in the classical (commutative) case (in which $\S(\H)$ is replaced by the set $\P_{+\infty}$ of
all probability distributions and the Shannon
entropy $S(\{x_i\})=-\sum_i x_i\log x_i$ is used instead of the von Neumann
entropy).

For an arbitrary sequence $\{h_i\}_{i=1}^{+\infty}$ of nonnegative
numbers and $h>0$ let
$\mathrm{g}(\{h_i\})=\inf\{\lambda>0\,|\,\sum_i\exp(-\lambda
h_i)<+\infty\}$ and
$$
\mathcal{K}_{\{h_i\},h}=\{\{x_i\}\in\P_{+\infty}\,|\,\textstyle\sum_i
h_i x_i \leq h\}
$$
be a closed convex subset of $\P_{+\infty}$. \smallskip

\begin{property}\label{c-case}
\emph{The Shannon entropy is continuous on a closed convex set
$\,\mathcal{A}\subset\P_{+\infty}$ if and only if
$\,\mathcal{A}\subseteq\mathcal{K}_{\{h_i\},h}$ for a particular sequence
$\{h_i\}$ of nonnegative
numbers such that $\,\mathrm{g}(\{h_i\})=0$ and $\,h>0$. }
\end{property}\smallskip

\textbf{Proof.} It suffices to prove the "only if " part, since the converse assertion follows from the
similar assertion for the von Neumann entropy (see the previous section).

Since the function $\{x_i\}\mapsto S(\{x_i\})$ is
finite on the closed convex set $\mathcal{A}$, it is bounded on this
set and the classical analog of Corollary 5 in \cite{Sh-4} shows
that the set $\mathcal{A}$ is compact. By Dini's lemma  the
condition $\|\{x_i\}\|_1=1$ and the continuity of the function
$\{x_i\}\mapsto S(\{x_i\})$ imply uniform convergence of the series
$\sum_i x_i$ and $\sum_i x_i(-\log x_i)$ on the set $\mathcal{A}$.
Hence there exists a sequence $\{y_i\}$ of positive numbers tending
to $+\infty$ such that
\begin{equation}\label{sp}
    \sup_{\{x_i\}\in\mathcal{A}}\sum_i y_i x_i<+\infty,
    \qquad \sup_{\{x_i\}\in\mathcal{A}}\sum_i y_i x_i(-\log
    x_i)<+\infty.
\end{equation}

Let $\mathcal{B}$  be the image of the set $\mathcal{A}$ under the
map $\{x_i\}\mapsto\{y_i x_i\}$. It follows from (\ref{sp}) that
$\mathcal{B}$ is a convex bounded subset of the positive cone of the
space $\ell_1$ and that the extended Shannon entropy (defined by the
formula $S(\{x_i\})=\|\{x_i\}\|_1 S\left(\frac{\{x_i\}}{\|\{x_i\}\|_1}\right)$)
is bounded on the set $\mathcal{B}$. By the classical analog of
Lemma 2 in \cite{Sh-12} there exists a sequence $\{h_i\}$ of
nonnegative numbers such that $\mathrm{g}(\{h_i\})<+\infty$ and
$\sup_{\{x_i\}\in\mathcal{A}}\sum_i h_i y_i x_i<+\infty$. It is easy to see
that $\mathrm{g}(\{h_i y_i\})=0$. $\square$

\section{General remarks}

The nontrivial part of the conjecture stated at the end of Section 1 can
be formulated as follows.\smallskip

\textbf{Conjecture 1.} An arbitrary closed convex set
$\mathcal{A}\subset\S(\H)$, on which the von Neumann entropy is
continuous, is contained in the set $\mathcal{K}_{H,h}$ for
a particular positive operator $H$ with $g(H)=0$ and $h>0$.\smallskip

The following proposition shows validity of this assertion for
simple types of convex closed subsets of states.\smallskip

\begin{property}\label{g-case}
\emph{The assertion of Conjecture 1 holds for a closed convex set
$\,\mathcal{A}\subset\S(\H)$ in the following cases:}
\begin{enumerate}[(i)]
    \item \emph{the set $\mathcal{A}$ consists of commuting states;}
    \item \emph{$\mathcal{A}$ is a convex hull of a finite collection of states;}\footnote{The von Neumann entropy is continuous on the set $\mathcal{A}=\mathrm{co}(\{\rho_i\}_{i=1}^n)$ if and only if $H(\rho_i)<+\infty$ for all $i$ \cite[Proposition 9a]{Sh-4}.}
    \item \emph{$\mathcal{A}=\Phi(\mathcal{B})$, where $\mathcal{B}$ is a
    compact subset of $\,\T_{+}(\K)$ with $\dim\K<+\infty$ and $\,\Phi$ is a
    positive linear map from $\T(\K)$ into $\T(\H)$.}\footnote{We denote by $\T(\H)$ and by $\T_+(\H)$ the Banach space of trace class operators in $\H$ and the positive cone in this space correspondingly.}
\end{enumerate}
\end{property}

\textbf{Proof.} $\mathrm{(i)}$ This immediately follows from Proposition
\ref{c-case}.\smallskip

$\mathrm{(ii)}$ If $\mathcal{A}=\mathrm{co}(\{\rho_i\}_{i=1}^n)$
then $H(\rho_i\|\bar{\rho})<+\infty$, where
$\bar{\rho}=n^{-1}\sum_{i=1}^n\rho_i$.\footnote{$H(\cdot\|\cdot)$ is the relative entropy.} Thus validity of the assertion in Conjecture 1
in this case follows from the implication
$\mathrm{(ii)}\Rightarrow\mathrm{(i)}$ in
\cite[Proposition 4]{Sh-4} (with $\mathcal{A}=\{\rho_i\}_{i=1}^n$ and $\sigma =\bar{\rho}$).\smallskip

$\mathrm{(iii)}$ We may assume that the set $\mathcal{B}$ contains
a full rank operator $B_0$.

Since the function $\rho\mapsto H(\rho)$ is continuous on the closed convex set
$\mathcal{A}$ it is bounded on this set and hence Lemma 2 in
\cite{Sh-12} implies existence of a positive operator $H$ in
$\mathcal{H}$ such that $\mathrm{g}(H)<+\infty$ and
$\mathrm{Tr}H\rho\leq h$ for all  $\rho\in\mathcal{A}$ and some
$h>0$. Finiteness of $\mathrm{g}(H)$ shows that $H=\sum_{i=1}^{+\infty}h_{i}|i\rangle\langle i|$, where $\{|i\rangle\}_{i=1}^{+\infty}$ is an orthonormal basis in
$\mathcal{H}$. Since $\Tr H\Phi(B_0)\leq h$ and  $B_0\geq \lambda I_{\K}$ for some $\lambda>0$,
we have
$$
\Tr H\Phi(I_{\K})=\sum_{i=1}^{+\infty}h_{i}\langle
i|\Phi(I_{\K})|i\rangle=\mathrm{Tr}\left[\sum_{i=1}^{+\infty}h_{i}\Phi^{*}(|i\rangle\langle
i|)\right]< +\infty
$$
and hence the linear operator in the square bracket lies in
$\mathfrak{B}(\mathcal{K})$. Thus the function
$$
B\mapsto\mathrm{Tr}H\Phi(B)=\sum_{i=1}^{+\infty}h_{i}\langle
i|\Phi(B)|i\rangle=\mathrm{Tr}\left[\sum_{i=1}^{+\infty}h_{i}\Phi^{*}(|i\rangle\langle
i|)\right]B
$$
is continuous on the compact set $\mathcal{B}$. By repeating the
arguments from the proof of Theorem 1 in \cite{Sh-12} based on
Dini's lemma one can construct a positive operator $H'$ in $\H$ such that
$\,\mathrm{g}(H')=0\,$ and $\,\sup_{\rho\in\mathcal{A}}\mathrm{Tr}H'\rho<+\infty$. $\square$
\smallskip

Consider equivalent forms of Conjecture 1.
\smallskip

Let $\mathcal{A}$ be a subset of $\S(\H)$. By
Proposition 5E in \cite{Sh-4} the following statements are related by
the implication $\mathrm{(ii)}\Rightarrow\mathrm{(i)}$:
\begin{enumerate}[(i)]
    \item the von Neumann entropy is continuous on the set
    $\mathcal{A}$;
    \item there exists an orthonormal basis $\{|i\rangle\}$ of the space $\H$ such that the Shannon entropy is continuous on the set
    $\{\{\langle i|\rho|i\rangle\}\,|\,\rho\in\mathcal{A}\}$.
\end{enumerate}
Proposition \ref{c-case} shows that\emph{ Conjecture 1 holds if and only
if the above statements $\mathrm{(i)}$ and $\,\mathrm{(ii)}$ are
equivalent for any closed convex set $\mathcal{A}\subset\S(\H)$.}\smallskip

By Proposition 4 in \cite{Sh-4} Conjecture 1 is equivalent to the
following one.\smallskip

\textbf{Conjecture 2.} For an arbitrary closed convex set
$\mathcal{A}\subset\S(\H)$, on which the von Neumann entropy is
continuous, there exists a state $\sigma$ in $\mathfrak{S}(\mathcal{H})$
such that the function  $\rho\mapsto H(\rho\|\sigma)$ is continuous
and bounded on the set $\mathcal{A}$.\footnote{It is sufficient to require that the function  $\rho\mapsto H(\rho\|\sigma)$ is continuous
and bounded on any closed subset $\mathcal{A}_0\subset\mathcal{A}$ whose convex closure coincides with $\mathcal{A}$.}
\bigskip

I would be grateful for any comments concerning the above questions.

\end{document}